\documentclass[a4paper,11pt]{article}
\pdfoutput=1
\usepackage{jcappub}
\usepackage[T1]{fontenc}
\usepackage{natbib}
\usepackage{txfonts,makecell}
\usepackage{hhline}
\usepackage{hyperref}
\usepackage{cleveref}
\usepackage{array}
\crefname{equation}{eq.}{eqs.}
\newcounter{mysfig}
\renewcommand\themysfig{(\alph{mysfig})}
\makeatletter
\newcommand\Scaption[1]{%
\refstepcounter{mysfig}%
\vskip.5\abovecaptionskip
  \sbox\@tempboxa{\small\themysfig~#1}%
  \ifdim \wd\@tempboxa >\hsize
    \small\themysfig~#1\par
  \else
    \global \@minipagefalse
    \hb@xt@\hsize{\hfil\box\@tempboxa\hfil}%
  \fi
  \vskip\belowcaptionskip}
\makeatother
\newcommand{\diff}{\mathrm{d}}

\title{Isotropic vs. Anisotropic components of BAO data: a tool for model selection}

  \author[a,b,c]{Balakrishna S. Haridasu}
  \author[b,c]{Vladimir V. Lukovi\'{c}}
  \author[b,c]{Nicola Vittorio}

  \affiliation[a]{Gran Sasso Science Institute , Viale Francesco Crispi 7, I-67100 L'Aquila, Italy}
  \affiliation[b]{Dipartimento di Fisica, Universit\`{a} di Roma "Tor Vergata", Via della Ricerca Scientifica 1, I-00133, Roma, Italy}
  \affiliation[c]{Sezione INFN, Universit\`{a} di Roma "Tor Vergata", Via della Ricerca Scientifica 1, I-00133, Roma, Italy}

  \emailAdd{sandeep.haridasu@gssi.it}
  \emailAdd{vladimir.lukovic@roma2.infn.it}
  \emailAdd{nicola.vittorio@roma2.infn.it}

  \abstract{ We conduct a selective analysis of the isotropic ($D_V$) and anisotropic ($AP$) components of the most recent Baryon Acoustic Oscillations (BAO) data. We find that these components provide significantly different constraints and could provide strong diagnostics for model selection, also in view of more precise data to arrive. For instance, in the $\Lambda$CDM model we find a mild tension of $\sim 2 \sigma$ for the $\Omega_m$ estimates obtained using $D_V$ and $AP$ separately. Considering both $\Omega_k$ and $w$ as free parameters, we find that the concordance model is in tension with the best-fit values provided by the BAO data alone at 2.2$\sigma$. We complemented the BAO data with the Supernovae Ia (SNIa) and Observational \textit{Hubble} datasets to perform a joint analysis on the $\Lambda$CDM model and its standard extensions. By assuming $\Lambda$CDM scenario, we find that these data provide $H_0 = 69.4 \pm 1.7$ \text{km/s Mpc$^{-1} $} as the best-fit value for the present expansion rate. In the $k\Lambda$CDM scenario we find that the evidence for acceleration using the BAO data alone is more than $\sim 5.8\sigma$, which increases to $8.4 \sigma$ in our joint analysis.
  }

\begin{document}
\maketitle
\flushbottom

\section{Introduction}

The Type Ia Supernovae (SNIa) compilation has so far provided the best observational constraints on the cosmological models for the late-time acceleration. At low-redshifts $(z\lesssim 2)$, the SNIa \citep{Betoule14} constraints were very well complemented by the Observational Hubble parameter (OHD) \citep{Simon05, Moresco16a, Moresco12a} and the Baryon Acoustic Oscillations (BAO) \citep{Eisenstein05, Alam16}.


Cosmological constraints from these low-redshift data were shown to be in a good agreement with those derived from the very precise high-redshift observations of Cosmic Microwave Background (CMB) \citep{Ade16}. As a result, the $\Lambda$CDM model has clearly emerged as the ``concordance model'' \citep{Aubourg15, Farooq16, Wang17a}. However, several questions have been raised regarding a concordance value for the $H_0$ \citep{Riess16, Lukovic16, Bernal16}. It has been shown time and again that the $H_0$ value from the direct estimate has been in a possible tension with the indirect model-dependent estimate. The more recent direct estimate \cite{Riess16} provides $H_0 = (73.24 \pm 1.74)$ \text{km/s Mpc$^{-1} $} (hereafter R16)\footnote{This value has been further revised in \citep{Anderson17} and \cite{Riess18}. These newer revisions do not show any significant change in the $H_0$ estimates and so we remain with R16 estimate.}, while the Planck collaboration \citep{Collaboration16b} derived $H_0 = (66.93 \pm 0.62)$ \text{km/s Mpc$^{-1} $} (hereafter P16)\footnote{See last column of Table 8 in \citep{Collaboration16b}.}. A similar lower value of $H_0=66.98 \pm 1.18$ \text{km/s Mpc$^{-1} $}, has been estimated by including the primordial deuterium abundance to the other datasets \citep{Addison17}.

Alongside the tension in the value of $H_0$, the current accelerated state of the Universe has also been questioned in \cite{Nielsen15}, which led to a discussion regarding the SNIa analysis and the evidence for a late-time acceleration in \cite{Rubin16, Haridasu17, Tutusaus17, Dam17}. More recently, the BAO dataset has been further improved, owing to the precise measurements from the SDSS (DR12) galaxy survey. Alam et al. (2016) \cite{Alam16} were able to disentangle the degeneracies in the transverse and the radial components in the redshift range $0.35 < z < 0.7$, providing both the transverse comoving distance ($D_M(z)$) and Hubble parameter ($H(z)$). With this improvement, the current BAO data is able to provide stronger constraints among the other low-redshift observations. In fact, using the BAO dataset alone now gives a significant evidence ($\sim 6.5\sigma$) for acceleration \citep{Ata17, Alam16}.

{In this paper we primarily focus on the Baryon Acoustic Oscillations data and a comparison of different information contained in different observables. The BAO data provides information about the cosmic expansion history through the characteristic scale of the acoustic oscillations at the ``high-redshift'' recombination epoch which is imprinted in the galaxy clustering and the intergalactic absorption of the Lyman-$\alpha$ forest \cite{Eisenstein98a}. This in turn asserts the basis of structure formation in linear theory of gravitational instability, for this reason the imprinted acoustic scale in the matter correlation function is referred to as a \textit{Standard Ruler}. This characteristic length scale is observed as an acoustic peak in the correlation function of the matter distribution at $\sim 100  h^{-1} \textrm{Mpc}$, first significant detection of which was reported in \cite{Eisenstein05}. The measurement of the acoustic peak is essentially a geometric complement to the luminosity based observations such as supernovae (\cite{Riess98}). As is evident the acoustic scale depends only on the expansion history during the early universe and the final observables are obtained only as scaled quantities, with respect to $r_d$.}

{The current sensitivity to the measurements of galaxy clustering is able to discern the angular-diameter distance and the expansion rate \cite{Alam16} observables.} The earlier measurements of the BAO data were reported for a volume averaged angular diameter distance $D_{V}(z)$ \citep{Eisenstein05}, which is a one-dimensional isotropic measurement and does not provide complete information. In fact, the missing information is contained in the anisotropic component, usually termed as the Alcock-Paczynski parameter $AP(z)$ \citep{Alcock79}. {The application of the $AP(z)$ parameter to test cosmologies has been addressed in several earlier works such as \cite{Li14, Li15, Li16}, for the redshift dependence of the same. However, an analysis using the complete anisotropic information from $D_M(z)$ and $H(z)$ (equivalently $D_{V}(z)$ and $AP(z)$) incorporates the $AP$-test in itself \cite{Alam16}, in contrast to using $D_V$ alone.} As these two components carry different information from the same observations, using only one of them could lead to biased results and hence incorrect inferences. While the importance of using the anisotropic information is known for many years, in this paper we utilise the newer anisotropic BAO measurements to quantify the differences in the constraints from the isotropic $D_V(z)$ and anisotropic $AP(z)$ components that can in-turn be used as a substantial method for falsifying models. We complement the BAO data with SNIa and OHD data to obtain joint constraints on cosmological models and further comment on the issues of $H_0$ and acceleration. We also show the effects of considering the approximate formulae \citep{Aubourg15} for the sound horizon at the drag epoch, while testing for the dynamical nature of dark energy.


The paper is organised as follows. In \Cref{sec:Model} the models tested are briefly described. In \Cref{sec:data} we present the data utilised in this paper, together with a brief description of the method. In \Cref{sec:ana} we report the results of our analysis. Finally, in \Cref{sec:con} we summarise our findings and discuss our main conclusions.

\section{Models}
\label{sec:Model}
 In this section we briefly describe the concordance $\Lambda$CDM model and its standard extensions that we test and compare in our analysis. The Friedmann equation with all standard degrees of freedom at low-redshifts is given by,
 \begin{equation}
\label{eqn:HUE}
H(z)^{2} = {H_0}^2\left[ \Omega_{m}(1+z)^3  + \Omega_{k}(1+z)^2 +\Omega_{DE}f(z)\right],
 \end{equation}
where, $H_{0}$ is the present expansion rate, while $\Omega_{m}$, $ \Omega_{DE}$ and $ \Omega_k $ are the dimensionless density parameters of matter, dark energy (DE) and curvature, respectively. The dimensionless density parameters obey the cosmic sum rule of $\Omega_{m}+\Omega_{DE}+\Omega_{k}=1$. The general functional form $f(z)$ gives the evolution of the DE and can be written as,
\begin{equation}
    f(z) = \exp\left( 3 \int^{z}_0 \frac{1+w(\xi)}{1+\xi} \diff \xi\right),
\end{equation}
where $w(z)$ is the equation of state (EOS) parameter of the dark energy. Hereafter, a constant in redshift EOS parameter is simply represented as $w$ in contrast to $w(z)$. For the flat $\Lambda$CDM model, $\Omega_{m}  = 1-\Omega_{DE} $ and $w=-1$. We test the standard extensions of $\Lambda$CDM model, namely the $k\Lambda$CDM model with the constraint $\Omega_{m}  = 1-\Omega_{\Lambda}-\Omega_{k}$, and the flat $w$CDM model with $w$ as a free parameter. The second Friedmann equation, $ \ddot{a}/a = - 4\pi/3 G\sum_i\rho_i(1+3w_i)$, gives us insight into the necessary conditions to be satisfied for assessing the dynamics of expansion rate. The criteria for acceleration can be derived as: $\Omega_{m} \leq \Omega_{\Lambda}/2$ for $k\Lambda$CDM and $w \leq -1/(3 \Omega_{\Lambda})$ for $w$CDM. We can assess the evidence for acceleration by estimating the confidence level at which these criteria are satisfied. One can also derive the deceleration parameter as,
  \begin{equation}
  \label{eqn:DP}
  q(z) = -a \frac{\ddot{a}}{\dot{a}^{2}} \equiv (1+z)\frac{H'(z)}{H(z)}-1.
  \end{equation}
A negative value of the deceleration parameter today, $q(0)$, implies an expanding universe at an accelerated rate. In addition, one can derive $q_0 = q(0) = 3/2 \Omega_m - \Omega_{\Lambda}$ for $\Lambda$CDM and $k\Lambda$CDM models and $q_0 = 1/2 (1+3w(1-\Omega_m))$ for the $w$CDM model.

We also study two parameter extensions to $\Lambda$CDM, namely $kw$CDM and the $w_{0}w_{a}$CDM. In the former, model both the $w$ and $\Omega_k$ are treated as free parameters. The latter model is given by Taylor expanding the EOS parameter around $a \sim 1$, as prescribed by the so-called CPL parametrisation \citep{Chevallier01, Linder03},
\begin{equation}
w(z) = w_{0} + w_{a}\frac{z}{1+z}.
\end{equation}
The luminosity distance for all these models can be written as,
\begin{equation}
    \label{eqn:luDb}
D_{L}(z) = (1+z)\frac{c}{H_{0}\sqrt{|\Omega_{k}|}} S\left( \sqrt{|\Omega_{k}|} H_{0} \int_{0}^{z} \frac{\diff\xi}{H(\xi)}\right)
\end{equation}
where,
 \begin{equation}
 \label{eqn:luD}
 S(x) \equiv \left\{
        \begin{aligned}
          & \sin(x).  &&\text{for}\,\,\Omega_{k}<0 \\
          & x.  &&\text{for}\,\,\Omega_{k}=0  \\
          & \sinh(x).   &&\text{for}\,\,\Omega_{k}>0
        \end{aligned} \right.
 \end{equation}
 The theoretical distance modulus is defined as $\mu_{th} =  5\log[D_{L}(\textrm{Mpc})] + 25$. The comoving angular diameter distance $D_{M}(z)$ is related to $D_{L}(z)$ as,
 \begin{equation}
 D_{M}(z) = D_{L}(z)/(1+z),
 \end{equation}
 which are used in the modelling of BAO data. Alternatively, the other two useful observables in modelling the BAO data are: $D_{V}(z)/r_{d}$ \citep{Eisenstein05} and $ AP(z)$ \citep{Alcock79}. The volume averaged comoving angular diameter distance $D_V(z)$ is given by,
 \begin{equation}
     \label{eqn:dv}
 D_{V}(z) = \left[D_{M}^{2}(z)\frac{c z}{H(z)}\right]^{1/3} \,
 \end{equation}
 and, $r_{d}$ is the sound horizon at the drag epoch ($z_{d}$) given as,
   \begin{equation}
     r_d =\int_{z_d}^{\infty} \frac{c_s(z)}{H(z)}dz.
     \label{eq:r_d}
   \end{equation}
Recently, in \cite{Aubourg15}, a functional form (hereafter A15) for the estimation of $r_d$ has been given as
   \begin{equation}
       r_d = 55.154 \frac{\exp{(-72.3(\Omega_{\nu} h^2 + 6\times10^{-4})^2})}{(\Omega_b h^2)^{0.12807} (\Omega_m h^2 - \Omega_{\nu} h^2)^{0.25351}} \, ,
       \label{eqn:aub}
   \end{equation}
where $\Omega_b h^2$ and $\Omega_{\nu} h^2$ are the baryon and neutrino densities, respectively. This functional form has been shown to be accurate up to sub-percent level.

The Alcock-Paczynski parameter was primarily defined as a test for the cosmological constant in \cite{Alcock79} and is written as,
\begin{equation}
    AP(z) = D_{M}(z) \frac{H(z)}{c}.
\end{equation}
One can easily construct the observables $AP(z)$ and $D_V(z)$ if the measurements for $D_M(z)$ and $H(z)$ are available.
{ The AP(z) observable in particular is of interest in our current work as we compare the constraints on the background cosmology from $AP(z)$ and the well-known $D_V(z)$ observable. In obtaining these observables it is necessary to assume a background/fiducial cosmology to establish the distance-redshift relation. An incorrect (different from the true model) assumption of the fiducial cosmology can introduce a systematic error while estimating the anisotropy in the clustering of galaxies induced by the peculiar velocities (often called as redshift-space distortions). This introduces a further anisotropy, a degenerate effect called geometric distortions (see \cite{Marulli12, Li14} for a detailed discussion) in the clustering observations which when evaluated can help assess the background cosmology, more often called as the $AP(z)$-test. This test has the advantage of being independent of evolutionary effects \cite{Alcock79} and the current data is able to provide a suitable significance for the same. }

\section{Data}
\label{sec:data}
In this section we describe the data used and the different methods adopted to test models against the data. In this current work we use the low-redshift BAO, OHD, and SNIa data.
\subsection{Baryon acoustic oscillations}
The BAO data until recently have been presented for the observable $D_{V}(z)/r_{d}$ \citep{Eisenstein05}, owing to the lack of sufficient statistics to distinctly measure $D_M(z)$ and $H(z)$\footnote{{While the observables are presented at times with the scaling as $D_M(z)\left(r_d^{fid}/r_d\right)$ and $H(z)\left(r_d^{fid}/r_d\right)^{-1}$, we express them simply in terms of $D_{M}(z)/r_{d}$ and $H(z) r_{d}$ in which way they are independent of the different fiducial cosmologies assumed in different works (similar to \cite{Aubourg15}).}}. {In fact, this lack of sufficient statistics restricted the earlier measurements from modelling the geometric distortions (see \Cref{sec:Model}) better and hence the spherically-averaged $D_{V}(z)/r_{d}$ was the only available observable.} The $D_V(z)/r_d$ variable has by far been used to constrain cosmological parameters providing good agreement with the SNIa data \citep{Betoule14}. These constraints improve once a joint analysis is performed (e.g., \citep{Lukovic16}).

In this work we utilise the measurements for $D_{M}(z)/r_{d}$ and $H(z) r_{d}$ (hereafter $D_{M}\&H$) and conduct a selective analysis using the different observables that one can obtain using these measurements. In \cite{Alam16} measurements of $D_{M}\&H$ (see their Table 8) at three binned redshifts $z = 0.32, 0.57, 0.61$ (hereafter 3z)\footnotemark were reported using the galaxy clustering data from Sloan Digital Sky Survey(SDSS) III. Earlier, \cite{Delubac15} have presented the observation of the BAO feature at the binned redshift $z =2.34$ in the flux-correlation of the Lyman-$\alpha$ forest of high-redshift quasars. Finally, the cross-correlation of Lyman-$\alpha$ forest absorption with the quasars has yielded another measurement at $z = 2.36$ in \cite{Ribera14}. These data points have been updated in \cite{MasdesBourboux17} and \cite{Bautista17}, with improvements implemented in their analyses. The newer data points are now given at redshifts $z=2.33$ and $z=2.4$. Although, other $D_{V}$ measurements at $z= 0.106, 0.15, 1.52$ \citep{Beutler11, Ross15, Ata17} are available, we do not implement them in our analysis as the $AP$ component at their respective redshifts is unavailable.
%

In \cite{Alam16}, the results have also been presented in terms of $D_{V}/r_d$ and $AP$ (hereafter $D_{V}\&AP$) parameter space, {which are independent of the assumed fiducial cosmology as also quoted in \cite{Alam16}}. It was also shown that the covariance between the $D_{V}/r_d$ and $AP$ points is negligible. Note that, the $D_{V}\&AP$ can be derived from the  $D_{M}\&H$ measurements (see \Cref{eqn:dv}) and then they are equivalent for cosmological parameter estimation. We construct a Jacobian matrix, $J$, to propagate the covariance from the $D_{M}\&H$ to the isotropic and the anisotropic basis given by $D_{V}\&AP$ (see \Cref{eqn:jacobian}).
\begin{equation}
    \label{eqn:jacobian}
\Sigma_{D_{V}\&AP}=J^{T} \cdot \Sigma_{D_{M}\&H} \cdot J,
\end{equation}
where $J$ is defined as the partial derivatives of the final functional form with respect to the initial one. Our formalism of constructing the $D_{V}\&AP$ data points converges very well to the covariances already presented in \cite{Alam16}. We implement the same formalism to obtain the $D_{V}\&AP$ data points along with their respective correlations for the Lyman-$\alpha$ data points. The correlations between the $D_{M}$ and $H$ estimates are given to be 0.377 and 0.369 at z=2.33 and z=2.4, respectively \citep{MasdesBourboux17}. Similarly, we also implement the tomographic BAO data for the 9 redshift bins presented in \cite{Zhao16, Wang17} (hearafter 9z)\footnotemark[\value{footnote}]. The constraints obtained from the 9z data were shown to be more stringent than those using 3z in \cite{Wang17a}. For brevity's sake, we show the measurements of $D_V$ and $AP$ only for 3z and Lyman-$\alpha$ data in \Cref{tab:baodata1}, and we summarise in \Cref{tab:baodata2} the covariances for the Lyman-$\alpha$ data points in the $D_{V}\&AP$ basis. One can notice the high correlation of the $D_{V}$ and $AP$ points at their respective redshifts. The correlation is found to be $\sim$ 0.8 and $\sim$ 0.53 at redshifts of 2.33 and 2.4, respectively. Given these high correlations, one must always use the full covariance matrix to compare the constraints coming from $D_{M}\&H$ with those obtained using $D_{V}\&AP$. Needless to say, this is also true for the 9z data (see also Figure 20 of \cite{Zhao16}). Please note that throughout the paper we refer to 3z+Ly-$\alpha$ data for the observables, unless otherwise quoted as 9z+Ly-$\alpha$ data.


\footnotetext{All values of the mean, dispersion and covariances of $D_{M}\&H$ observables for the galaxy clustering BAO data are taken from \href{https://data.sdss.org/sas/dr12/boss/papers/clustering/}{https://data.sdss.org/sas/dr12/boss/papers/clustering/} . \cite{Alam16} also provides the covariances for 3z $D_{V}\&AP$, while \cite{Zhao16} provides the covariances only for 9z $D_M\&H$.}

\begin{table}[ht]
\begin{center}
\label{tab:baodata1}
\footnotesize
\vspace{0.2in}
\begin{tabular}{|c|c|c|c|}
\hline
 $z$ &$D_{V}/r_{d}$ & $AP$ & Reference  \\
\hline
0.38 &  9.995 $\pm$ 0.111 & 0.413 $\pm$ 0.013 & I \\
\hline
0.51 & 12.700 $\pm$ 0.129 & 0.597 $\pm$ 0.017 & I \\
\hline
0.61 &  14.482 $\pm$ 0.149 & 0.741 $\pm$ 0.021 & I \\
\hline
2.33 & 31.123 $\pm$ 1.087 & 4.164 $\pm$ 0.317 & II \\
\hline
2.4 &  30.206 $\pm$ 0.892 & 3.962 $\pm$ 0.288 & III\\
\hline
\end{tabular}
\caption{BAO data in $D_{V}\&AP$ formalism. The reference I corresponds to \cite{Alam16}, which provides both the data and the covariances.
    We construct the Lyman-$\alpha$ $D_{V}\&AP$ data using the measurements $D_{M}\&H$ from references II \citep{Bautista17} and III \citep{MasdesBourboux17}.}
\end{center}
\end{table}

{\renewcommand{\arraystretch}{1.45}%
\begin{table}[ht]
\begin{center}
\label{tab:baodata2}
\footnotesize
\vspace{0.2in}
\begin{tabular}{|c|cccc|}
\hline
  Measurement &  \multicolumn{4}{c|}{$c_{ij}$} \\
\hline
$D_V(2.33)$  & 1.18205 & 0. & 0.25571 & 0. \\

$D_V(2.4)$ & 0. & 0.79666 & 0. & 0.14525 \\

$AP(2.33)$  & 0.25571 & 0. & 0.10061 & 0. \\

$AP(2.4) $  & 0. & 0.14525 & 0. & 0.08271 \\
\hline
\end{tabular}
\caption{For completeness we show here the total covarince matrix ($C_{ij}$) in $D_{V}\&AP$ basis obtained using \Cref{eqn:jacobian} for the two Lyman-$\alpha$ points at z = 2.33 and z = 2.4.}
\end{center}
\end{table}
}

The likelihood for the correlated BAO data is implemented as,
\begin{equation}
    \label{eqn:Slikelihood}
 {\cal L}_{\textrm {BAO}}\propto\exp\left[(Y_{data}-Y_{th}).\Sigma_{BAO}^{-1}.(Y_{data}-Y_{th})^{T}\right],
\end{equation}
where $Y_{data}$ is the data described in \Cref{tab:baodata1} and the $Y_{th}$ is the theoretical model evaluated at the respective redshifts, while $\Sigma_{BAO}$ denotes the total covariance matrix of the BAO data for either of the observables ($D_V$, $AP$, $D_{V}\&AP$ and $D_{M}\&H$). As the $AP$ component is independent of $r_d$, unlike $D_V/r_d$, we use $r_d \times H_0$ (hereafter $H_0r_d$) as a free parameter, instead of using A15, to compare the individual constraints obtained from these components. However, we also implement the A15 functional form to compare the results obtained from our main analysis. For this purpose, we use \Cref{eqn:aub} with $\Omega_b h^2 = 0.0217$ \citep{Riemer-Sorensen17} and $\Omega_{\nu} h^2 = 6.42\times10^{-4}$ \citep{Ade16}.

\subsection{Cosmic chronometers and supernovae Ia}
The measurements of the expansion rate have been estimated using the differential age method suggested in \cite{Jimenez02}, which considers pairs of passively evolving massive red galaxies at similar redshifts to obtain $\diff z/\diff t$. {The expansion rate of the universe is obtained from these ``cosmic chronometers'', where the relative ages of the passive galaxies are inferred as a \textit{Standard(-izable) Clocks} and hence one can obtain $H(z) = -1/(1+z)\diff z/\diff t$.} We use the compilation of 31 uncorrelated data points taken from \cite{Simon05, Stern10, Moresco12b, Moresco16a, Moresco15, Zhang14, Ratsimbazafy17}. A similar compilation was also implemented in \cite{Farooq16}, but also considering additional $H(z)$ measurements from BAO data.

We implement a simple likelihood function assuming all the data are uncorrelated as,
\begin{equation}
{\cal L}_{\textrm {OHD}}\propto\exp\left[-\dfrac{1}{2}\sum_{i=1}^{31} \left(\frac{ H_i-H(z_i)}{\sigma_{H_i}}\right)^2\right]
,\end{equation}
where $\sigma_{H_i}$ is the gaussian error on the measured value of $H_i$.

We use the JLA supernovae compilation {$\sim 740$} SNe, with the standard $\chi^{2}$ analysis as is elaborated in \cite{Betoule14}. {The SNIa provide a luminosity based distance-redshift observations and are deemed as \textit{Standard Candles}, as each supernovae is expected to have the same peak luminosity (i.e., absolute magnitude ($M_B$)). However, it is now known that the observed SNIa are not perfect standard candles and one needs to implement corrections to standardise them (see e.g., \cite{Tripp98, Tripp99, Guy07, Betoule14}).} The JLA supernova compilation implements a correction to the absolute magnitude through the empirical relation,
  \begin{equation}
    M_B^{\textrm {corr}}=M_B-\alpha s + \beta c + \Delta M \Theta \left[\log\left(\frac{M^{Host}}{M_{\odot}}\right)-10\right],
  \end{equation}
{ where $M_B$ is the absolute magnitude, $s$, $c$ and $\Delta M$ are the stretch, colour and host galaxy mass corrections for the absolute magnitude. Here $\Theta$ represents the Heaviside function and $M^{Host}$ is the mass of the host galaxy. It is worthwhile noting that the SNIa compilation retains an additional unaccounted scatter in the apparent magnitude ($m_B$) even after the corrections are implemented, this is also known as the \textit{Hubble residual}. Therefore,  the likelihood for the SNIa data has been implemented by model-independently evaluating this intrinsic scatter ($\sigma_{int} \sim 0.106$) which is accounted for in the total covariance matrix ($Cov$) of the SNIa compilation.} The observed distance modulus $\mu_{obs}$ can be written as,
  \begin{equation}
      \mu_{obs} = m_B - M_B^{\textrm {corr}} + \alpha s -\beta c -  \Delta M \Theta \left[\log\left(\frac{M^{Host}}{M_{\odot}}\right)-10\right],
  \end{equation}
 The likelihood for the SNIa data can be written as,
 \begin{equation}
     \label{eqn:SNlikelihood}
 {\cal L}_{\textrm {SN}}\propto\exp\left[\frac{(\mu_{obs}-\mu_{th})^{T} \cdot Cov^{-1} \cdot (\mu_{obs}-\mu_{th})}{|Cov|}\right]
 ,\end{equation}
where $Cov$ denotes the total covariance matrix. {In this analysis $\alpha$, $\beta$ and $M_B$ are treated as nuisance parameters. As can be seen from \Cref{eqn:luDb} and \Cref{eqn:SNlikelihood}, the parameters $M_B$ and $H_0$ remain degenerate in the SNIa Likelihood. }


\subsection{Joint analysis and statistical inference}

Finally, the joint analysis is performed by combining the likelihoods for these independent datasets. The likelihood for the joint analysis is given as,
\begin{equation}
    \label{eqn:Jlikelihood}
    {\cal L}_{\textrm{TOT}} = {\cal L}_{\textrm {BAO}}{\cal L}_{\textrm {OHD}}{\cal L}_{\textrm {SN}}.
\end{equation}

{As is known the present rate of expansion $H_0$ is degenerate with absolute magnitude ($M_B$) and sound horizon ($r_d$) in the SN and BAO data, respectively. To this aid, the OHD data which provides the information of expansion rate alone, plays a very important role in our joint analysis as it is capable of discerning the degeneracy between the Hubble parameter $H_{0}$ and $M_{B}$ for supernova data, and the degeneracy between $H_{0}$ and $r_{d}$ in case of BAO data. Hence, in our joint analysis we are able to constrain $H_0$, which is complemented by the ability of SN and BAO data to constrain the energy densities better than the OHD data.}
{We implement a frequentist approach to optimise the likelihood (\Cref{eqn:Slikelihood} or \Cref{eqn:Jlikelihood}) to obtain the best-fit parameters and then proceed to estimate the 2-dimensional confidence levels and uncertainties of the individual parameters by optimising over the rest of the parameters. We have also verified that the results remain unaltered even if a Bayesian analysis is performed. Note that we show only the contours of the parameters relevant for our discussion, please refer to the tables provided for the constraints on all other parameters. }

We use the Akaike information criteria (AIC) \citep{Akaike74} and Bayesian information (BIC) \citep{Schwarz78} for model selection. The AIC and AICc (corrected for number of data points) with a second-order correction term, are written as,
\begin{align}
\text{AIC} &= -2\log{\cal L}^{max} + 2 N_p, \\
\text{AICc} &= -2\log{\cal L}^{max} + 2 N_p + \frac{2N_p(N_p+1)}{N_d-N_p-1},
\end{align}
where $N_p$ is the number of parameters and $N_d$ is the number of data points. For large $N_d$, the AICc value tends to AIC, while for less $N_d$ (e.g., BAO data) it penalises the model with more parameters strongly. Similarly, the BIC can be defined as,
\begin{equation}
\text{BIC} = -2\log{\cal L}^{max} + N_p\log(N_{d}),
\end{equation}
{Note that both these criteria are made in order to prefer a model with fewer parameters, when they are able to fit the data equally well.} The model preference is estimated by evaluating $\Delta$AICc ($\Delta$BIC) as a difference in the AICc value of the model in question to the reference model. A positive value of $\Delta$AICc or $\Delta$BIC indicates that the reference model is preferred over the model in comparison. {The AIC statistic essentially evaluates the amount of information lost from one model to another, which are used to describe the data. Similarly, BIC provides selection based on posterior probability of the models in comparison. As mentioned above, these criteria provide a more suitable way to evaluate model selection as they penalise the models for the additional number of parameters compared to a simple comparison of $\log{\cal L}^{max}$, which is bound to prefer a model a built with more number of parameters. Also, note that $\Delta$BIC penalises the models with more parameters more strictly than $\Delta$AICc. They have been implemented on numerous occasions in the cosmological context (see e.g., \cite{Verde10, Bernal16, Shafer15}), ever since their introduction in \cite{Liddle04}.}

{We also implement the newly proposed Index of Inconsistency (IOI) \citep{Lin17, Lin17a} to measure the inconsistency between constraints obtained form two or more different datasets. The IOI for a one parameter distribution can be defined as,
\begin{equation}
    \textrm{IOI} = \frac{\left(\mu_1 -\mu_2\right)^2}{2(\sigma^2_1+\sigma^2_2)},
\end{equation}
where $\mu_i$ and $\sigma_i$ are the mean and uncertainty in the estimates obtained using dataset $i$, respectively. This estimate of inconsistency can be related to the standard confidence level tension as $\sqrt{2 \textrm{IOI}} $ $ \sigma$ for two one parameter distributions. The IOI can be generalised to more than two estimates of Gaussian likelihoods/distributions as,
\begin{equation}
    \label{eqn:IOIN}
    \textrm{IOI} \equiv \frac{1}{N} \sum_{i}^{N} \Delta\log({\cal L}_{i}^{max})(\mu) = \frac{1}{N}\left[\sum_{i}^{N} \frac{\mu_i^{2}}{\sigma_{i}^{2}} - \frac{\left(\sum_{i}^{N}\mu_{i}\sigma_{i}^{-2}\right)^{2}}{\left(\sum_{i}^{N} \sigma_{i}^{-2}\right)} \right],
\end{equation}
where $N$ is the number of independent estimates, $\mu_{i}$ and $\sigma_{i}$ have their usual meanings (see Eq. 11-13 of \cite{Lin17}). The term $\frac{1}{N} \sum_{i}^{N} \Delta\log({\cal L}_{i}^{max})(\mu)$, describes the averaged difficulty of the individual estimates to support their joint mean, where ${\cal L}_{i}^{max}$ is the maximised likelihood of $i^{th}$ dataset. In our analysis we implement the IOI for the case of $N=3$ in \Cref{eqn:IOIN}, when required.}

\section{Results and Discussion}
\label{sec:ana}
In this section we present the results obtained from our analysis for BAO data alone and then for the joint analysis described in the earlier sections. We also comment on the statistical evidence for a late-time acceleration phase.
\subsection{Results from the analysis of BAO data alone}
As mentioned in the earlier sections we have used the different observables ($D_V$, $AP$, $D_{V}\&AP$ and $D_{M}\&H$) taken from the BAO data to constrain the standard model and its extensions. In particular, we test for the agreement $D_V$ and $AP$ components. In \Cref{tab:baofitlcdm}, \Cref{tab:baofitklcdm} and \Cref{tab:baofitwcdm} we show the best-fit parameters to $ \Lambda $CDM, $k\Lambda$CDM and $w$CDM models using each of the four observables. Except for the $k\Lambda$CDM model, the $AP$ data gives a lower value of $\Omega_{m}$ for all the standard extensions considered, while the $D_V$ measurements give higher values of the same with similar ability to constrain the mean.

\begin{figure}[ht]
\includegraphics[width=0.45\textwidth]{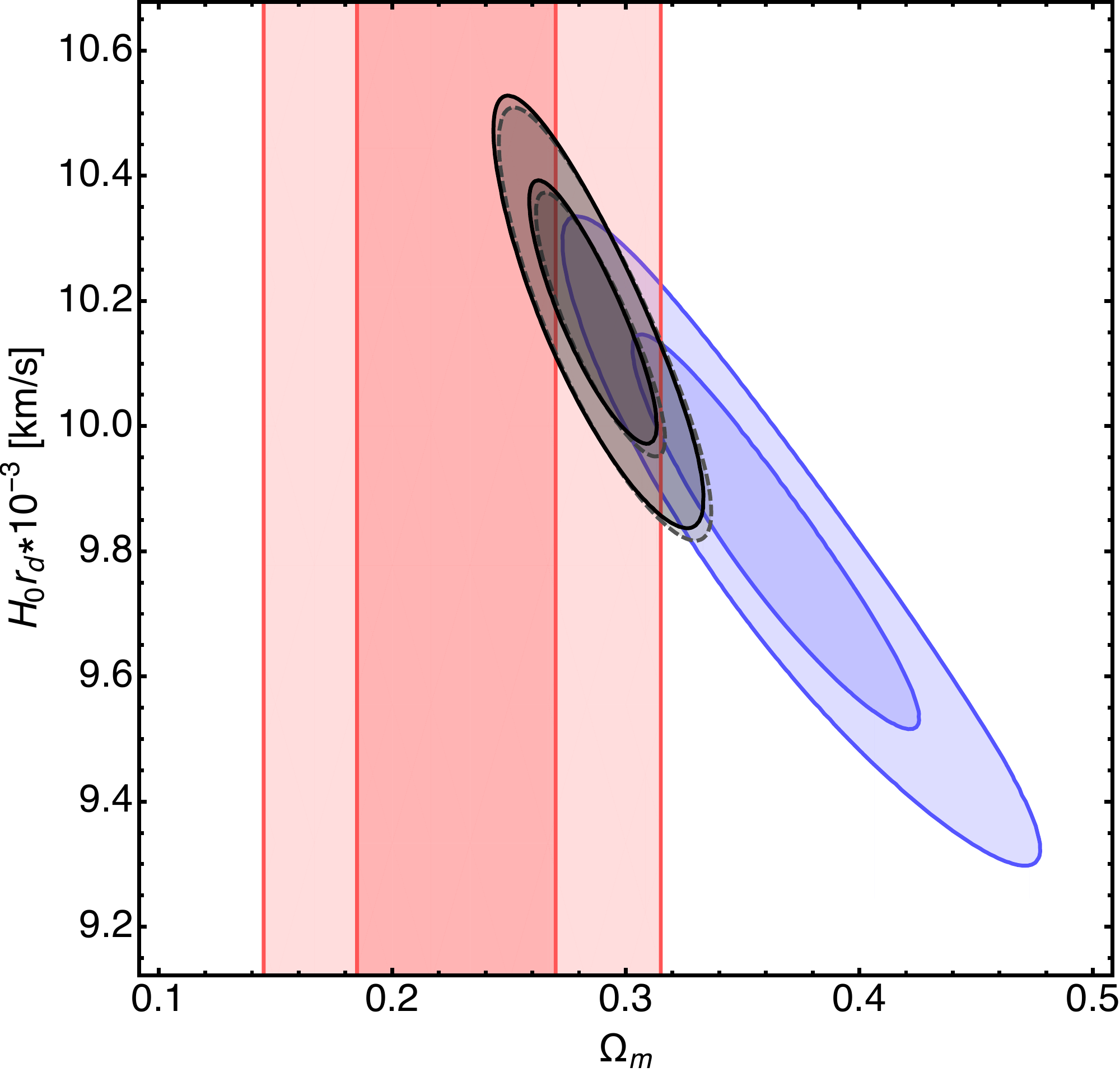}
\hspace{0.2in}
\includegraphics[width=0.45\textwidth]{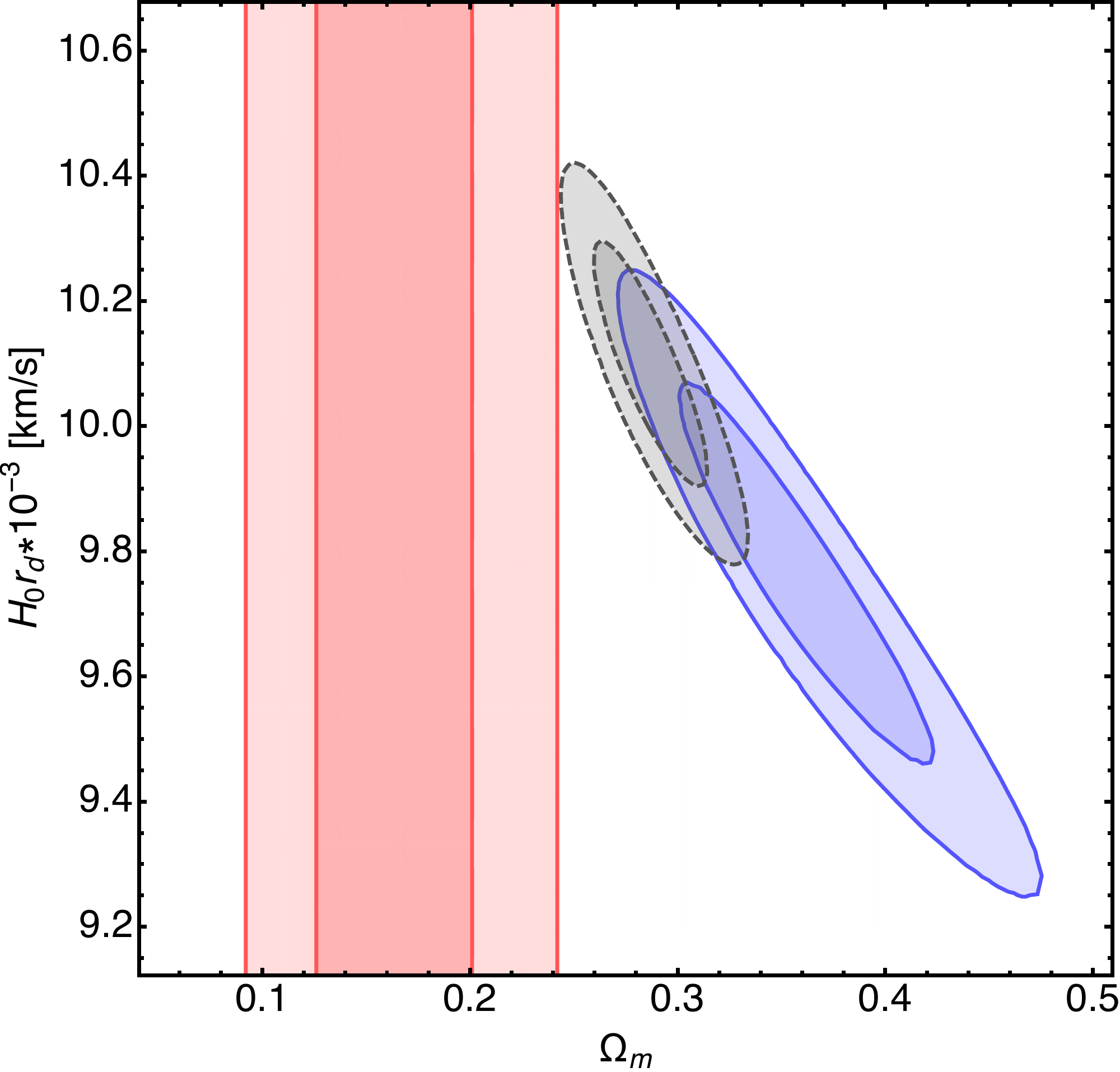}
\caption{We show the 68\% and 95\% confidence level contours for 3z+Ly-$\alpha$ (left) and 9z+Ly-$\alpha$ (right) in $\Lambda$CDM scenario. The red, blue and black contours correspond to $AP$, $D_{V}$ and $D_{V}\&AP$ methods, respectively. The gray dashed contours in both the panels shows the constrains using $D_{M}\&H$ formalism.}
\label{fig:con1}
\end{figure}

In the $\Lambda$CDM scenario, the $\Omega_m$ estimates obtained using $D_V$ and $AP$ as independent datasets show a mild tension\footnote{As the errors are asymmetric, we assume a gaussian error using the higher among them to estimate the tension.} at $2.1 \sigma$. However, this tension might be overestimated if the two values of $\Omega_m$ are anti-correlated\footnote{It is easily verified that for positively correlated $D_V$ and $AP$ data, the two values of $\Omega_m$ will be anti-correlated.}. Therefore, we estimate a least possible tension of $1.5 \sigma$ by assuming that they are completely anti-correlated. Using the 9z+Ly-$\alpha$ data we find the $\Omega_m$ values from $D_V$ and $AP$ to be $0.356^{+0.043}_{-0.037}$ and $0.159^{+0.041}_{-0.034}$, respectively. Here the discrepancy is even more pronounced and they agree only at $3.3 \sigma$ considering no correlation and a least possible tension of $2.3 \sigma$ (see \Cref{fig:con1}). As expected, we find that using the BAO data in either the $D_{M}\&H$ or $D_{V}\&AP$ formalism yield very consistent results suggesting that there is no loss of information and that these two parameter spaces are equivalent (see \Cref{tab:baofitlcdm}).

{\renewcommand{\arraystretch}{1.45}%
\begin{table}[ht]
\begin{center}
\label{tab:baofitlcdm}
\begin{tabular}{|ccc|}
\hline
 Data & $\Omega_{m}$ & $H_{0}r_{d}[\text{km/s}] $  \\
\hline
$AP$ & $0.225^{+0.045}_{-0.040}$ &  -    \\

$D_{V}$ & $0.358^{+0.043}_{-0.038}$ & $9840^{+204}_{-212}$  \\

$D_{V}\&AP$ & $0.285^{+0.019}_{-0.017}$ & 10182 $\pm$ 139  \\

$D_{M}\&H$ &  $0.288^{+0.019}_{-0.018}$ & 10162 $\pm$ 139 \\
\hline
\end{tabular}
\caption{Fit parameters for the four different observables in $\Lambda$CDM model.}
\end{center}
\end{table}
}

{\renewcommand{\arraystretch}{1.45}%
    \begin{table}[ht]
        \begin{center}
            \label{tab:baofitklcdm}
            \begin{tabular}{|cccc|}
                \hline
                Data &$\Omega_{m}$ & $\Omega_{\Lambda}$ & $H_{0}r_{d}[\text{km/s}]$  \\
                \hline
                $AP$ & $1.748^{+0.561}_{-0.521}$ & $1.720^{+0.306}_{-0.301}$&-    \\
                
                $D_{V}$ & $0.358^{+0.045}_{-0.040}$ & $0.537^{+0.224}_{-0.306}$ & $9632^{+504}_{-550}$  \\
                
                $D_{V}\&AP$ &  0.300  $\pm$ 0.025 & $0.786^{+0.079}_{-0.085}$ & $10303^{+199}_{-199}$ \\
                
                $D_{M}\&H$ &  0.302  $\pm$ 0.024 & $0.783^{+0.081}_{-0.087}$& 10285 $\pm$ 202 \\
                \hline
            \end{tabular}
            \caption{Best-fit parameters in the four different formalisms for $k\Lambda$CDM model.}
        \end{center}
    \end{table}
}

\begin{figure}[ht]
\includegraphics[width=0.45\textwidth]{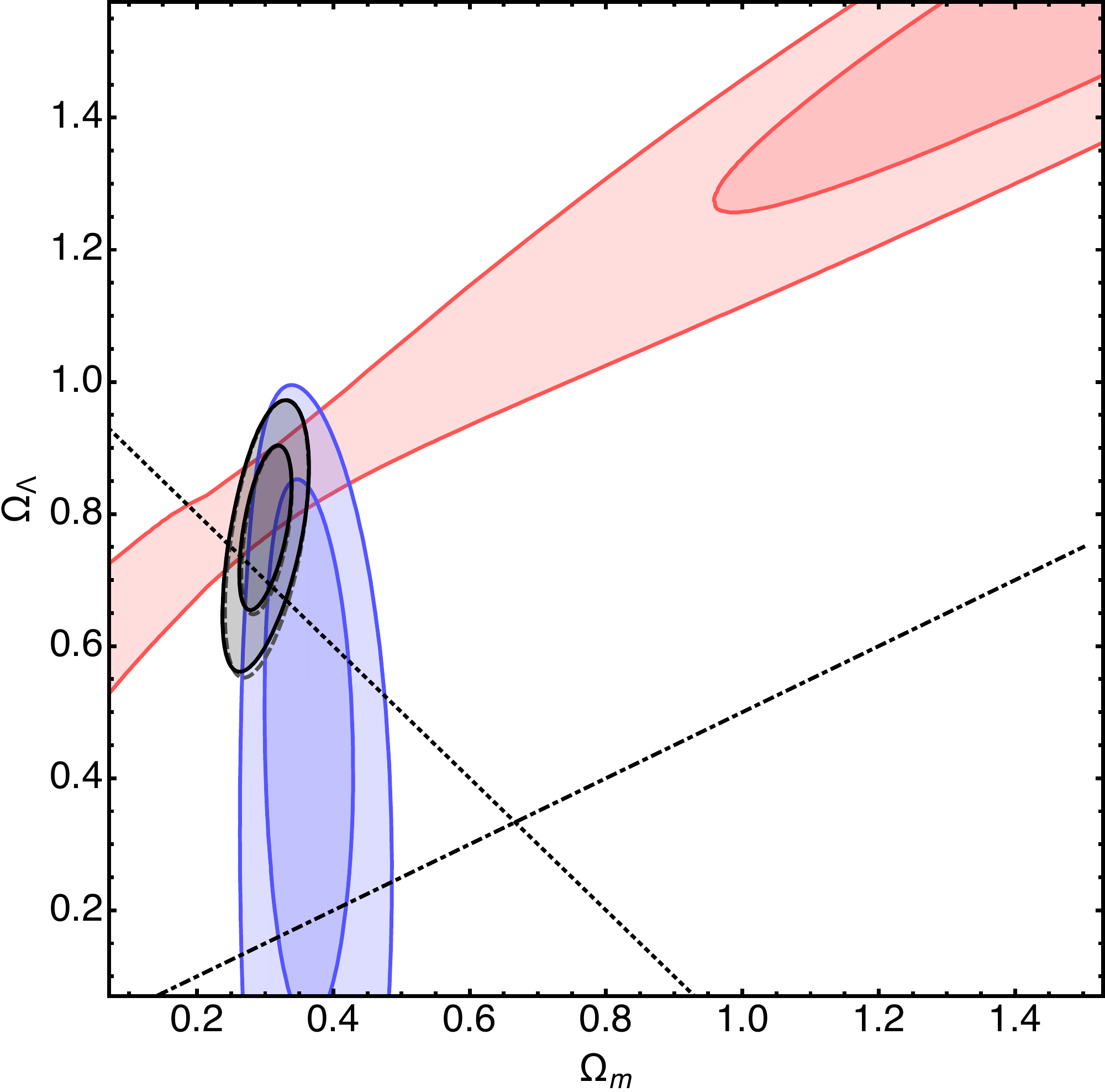}
\hspace{0.2in}
\includegraphics[width=0.45\textwidth]{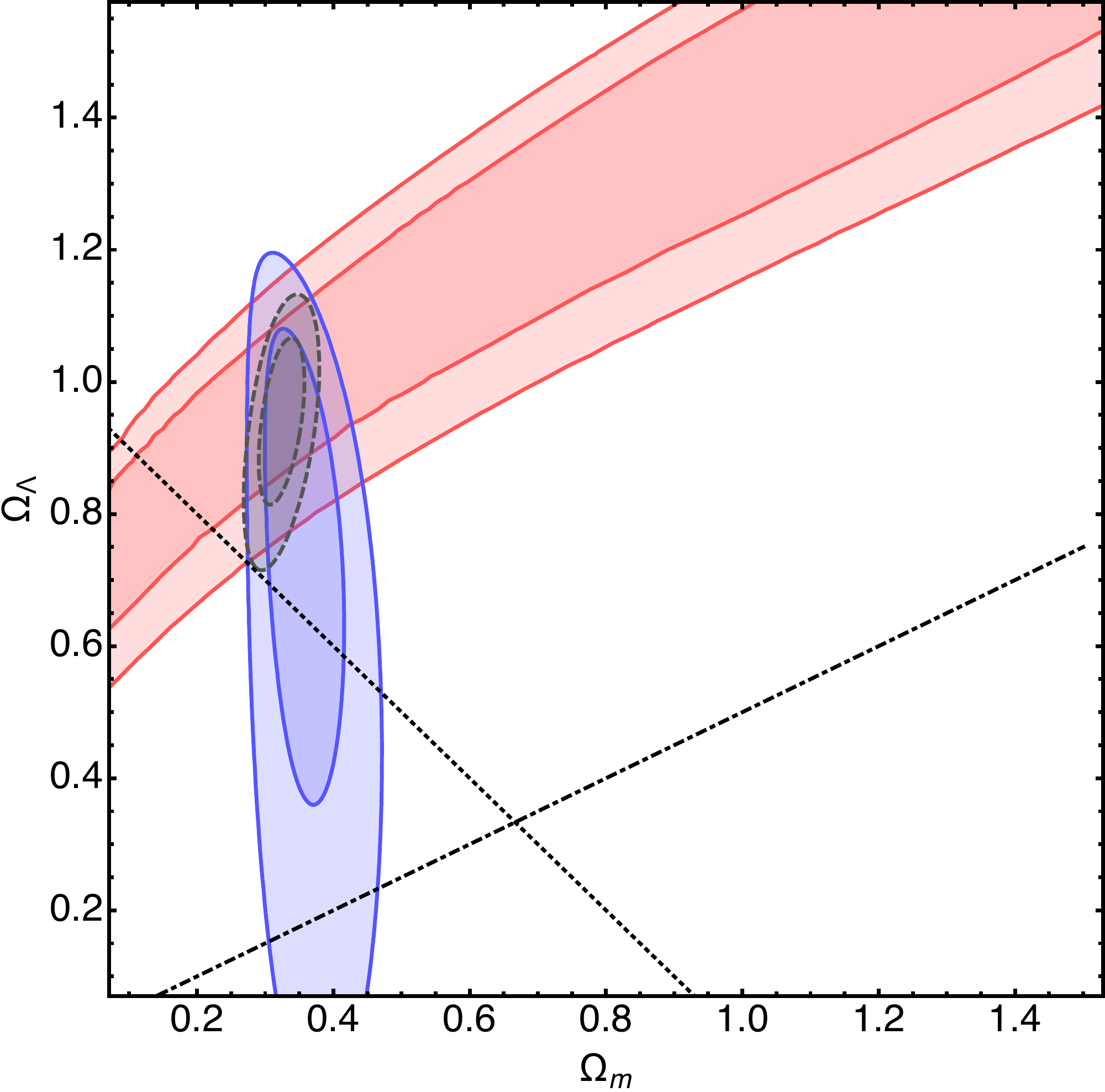}
\caption{We show the 68\% and 95\% confidence level contours for 3z+Ly-$\alpha$ (left) and 9z+Ly-$\alpha$ (right) in $k\Lambda$CDM scenario. The red, blue and black contours correspond to $AP$, $D_{V}$ and $D_{V}\&AP$ methods, respectively. The dashed contours in both the panels shows the constrains using $D_{M}\&H(z)$ formalism. The dotted and the dot-dashed lines identify the flat $\Lambda$CDM and the transition between the accelerated and non-accelerated regimes, respectively.}
\label{fig:con2}
\end{figure}

\begin{figure}[ht]
    \centering
\includegraphics[width=0.45\textwidth]{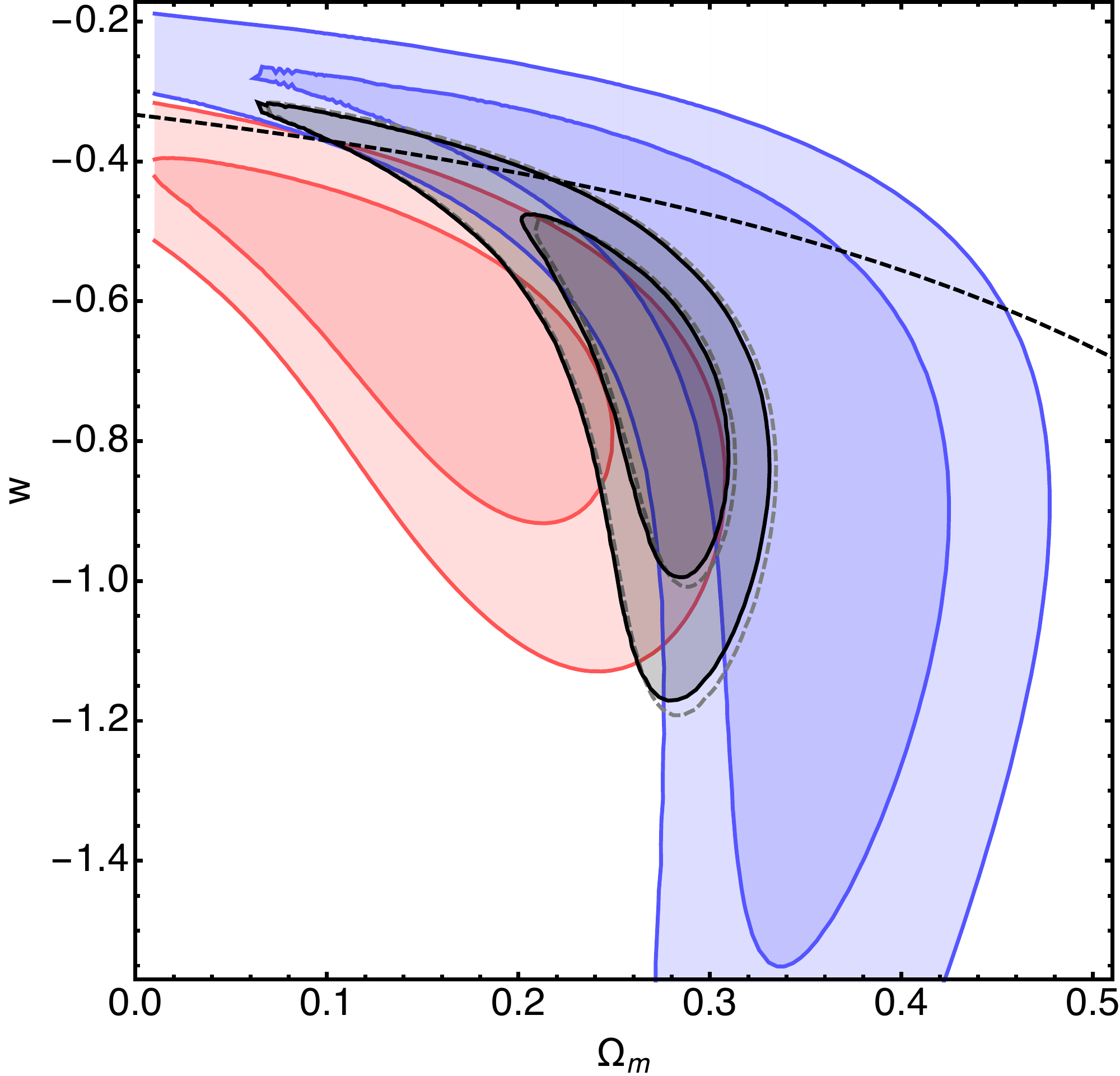}
\caption{The red, blue and black contours correspond to $AP$, $D_{V}$ and $D_{V}\&AP$ methods, respectively. We show here the 68\% and 95\% confidence level contours. The Dashed contours in grey show the constrains using $D_{M}\&H(z)$ method. The region below (above) the dashed line corresponds to the accelerating (non-accelerating) regime.}
\label{fig:con3}
\end{figure}

In the $k\Lambda$CDM model we find that using $AP$ data alone results in unexpected high values of $\Omega_m$ and $\Omega_{\Lambda}$ (see \Cref{fig:con2}). Also, a consistent estimate for the flatness in the $k\Lambda$CDM model can be obtained only when both $D_{V}$ and $AP$ components are used together (see left panel of \Cref{fig:con2}). However, using 9z+Ly-$\alpha$ data we find that the flat $\Lambda$CDM model is disfavoured at $\sim 2\sigma$ (see right panel of \Cref{fig:con2}). In the $w$CDM model, we see that the ability of $AP$ measurements to constrain the value of $w$ is better than that of $D_V$ measurements. Hence, the inclusion of the $AP$ measurements is crucial to have reliable inferences, e.g., on the phantom scenario (see \Cref{fig:con3}). We want to stress that in all the models considered here the discrepancies between the $D_V$ and $AP$ constraints remain consistent from 3z to 9z data.

{\renewcommand{\arraystretch}{1.45}%
\begin{table}[ht]
\begin{center}
\label{tab:baofitwcdm}
\begin{tabular}{|cccc|}
\hline
 Data &$\Omega_{m}$ & $ w $ & $H_{0}r_{d}[\textrm{km/s}]$  \\
\hline
$AP$ & $0.157^{+0.062}_{-0.082}$ & $-0.646^{+0.166}_{-0.174}$& -    \\

$D_{V}$ & $0.353^{+0.046}_{-0.104}$ & $-0.822^{+0.387}_{-0.443}$ & $9570^{+727}_{-551}$   \\

$D_{V}\&AP$ & $0.275^{+0.023}_{-0.035}$ & $-0.736^{+0.166}_{-0.169}$ & $9692^{+334}_{-304}$\\

$D_{M}\&H$ &  $0.278^{+0.024}_{-0.036}$ & $-0.741^{+0.171}_{-0.174}$ & $9690^{+340}_{-308}$ \\
\hline
\end{tabular}
\caption{Best-fit parameters to the BAO data in the four different methods for $w$CDM model.}
\end{center}
\end{table}
}

The constraints from BAO data for the $w_0w_a$CDM and the $kw$CDM models are discussed in the next subsection, alongside the joint analysis. Inclusion of other three BAO $D_V$-only points at $z= 0.106, 0.15, 1.52$ affects the constraints from the BAO data mildly. For instance, in the $\Lambda$CDM model we obtain $\Omega_m = 0.293^{+0.18}_{-0.17}$, showing clearly the minimal effect that these data have on constraining the model. We now proceed with the joint analysis without including these three data points.

\subsection{Joint analysis and model selection}

In this section we present our constraints obtained from a joint analysis of the most recent $D_M\&H$ + SN + OHD data. Our analysis (similar to \cite{Moresco16}) differ from what is already present in the literature, as the BAO data are usually complemented with the CMB data. Here we don't consider the CMB data in order to test the consistency between the ``low-redshift'' and ``high-redshift'' constraints. Moreover, in order to estimate $H_0$, we rely on a joint analysis which includes the OHD dataset, more often not considered in this kind of analysis. In \cref{tab:joint} we show our findings. Note that the estimates for $H_0$ and $r_d$ are consistent among all the models considered here. Note also the low value of $\Omega_m$ obtained for the $w_0w_a$CDM model.

{We find the best-fit estimate of $H_0$ for the $\Lambda$CDM model to be $H_0 = 69.41 \pm 1.76$ \textrm{km/s Mpc$^{-1} $}. Note that this value is now in the middle of P16 ($H_0 = 66.93 \pm 0.62$ \textrm{km/s Mpc$^{-1} $}) and R16 ($H_0 = 73.24 \pm 1.74$ \textrm{km/s Mpc$^{-1} $}) estimates, and is consistent with both at $1.33\sigma$ and $1.55\sigma $, respectively. To quantify the inconsistency between the three independent $H_0$ estimates -from this work, P16 and R16- we utilise the IOI estimate alone and obtain a total IOI = 4.19. This inconsistency is lower than the inconsistency of IOI = 5.83 (also quoted in \cite{Lin17a}) between P16 and R16, which corresponds to a standard tension of 3.4$\sigma$. The total inconsistency amongst these three estimates reduces to $\textrm{IOI} = 0.88$ and $\textrm{IOI} = 1.19$ when R16 and P16 are excluded, respectively. Consequently, these inconsistencies do not indicate any preference for either of the three estimates, unlike in \cite{Lin17a}, where the R16 estimate was quoted as a possible outlier.}

{\renewcommand{\arraystretch}{1.45}%
    \begin{table}[ht]
        \begin{center}
            \label{tab:joint}
            \resizebox{\columnwidth}{!}{%
                \begin{tabular}{|ccccccc|}
                    \hline
                    Model &$\Omega_{m}$ &$H_{0}$[\text{km/s Mpc$^{-1} $}] & $ w_{0}/w $ & $w_{a}$&$\Omega_{\Lambda}$ & $r_d$[Mpc] \\
                    \hline
                    
                    $\Lambda$CDM & $0.292^{+0.016}_{-0.015}$  & $69.41 \pm 1.76$ & - & - & -  & $146.0^{+3.6}_{-3.4}$\\
                    
                    $k\Lambda$CDM & $0.296 \pm 0.024$  & $69.62^{+2.00}_{-1.98}$ & - & - & $0.722^{+0.064}_{-0.067}$ & $145.9^{+3.7}_{-3.5}$\\
                    
                    $w$CDM & $0.285 \pm 0.018$  & $68.61^{+1.93}_{-1.91}$ & $-0.921^{+0.08}_{-0.081}$& - & - & $146.2^{+3.6}_{-3.4}$\\
                    
                    $w_{0}w_{a}$CDM & $0.195^{+0.084}_{-0.230}$  & $68.75^{+1.95}_{-1.92}$& $-0.902^{+0.222}_{-0.125}$&  $0.838^{+0.217}_{-0.655}$ & - &$146.5^{+3.6}_{-3.5}$ \\
                    
                    $kw$CDM &$0.311 \pm 0.025$ & $69.27^{+2.00}_{-1.97}$& $-0.828^{+0.075}_{-0.089}$& - &$0.872 \pm 0.107 $ & $144.9^{+3.7}_{-3.5}$\\
                    
                    \hline
                \end{tabular}%
            }
            \caption{Best-fit estimates and parameter constraints using the $D_M\&H$+SN+OHD data.}
        \end{center}
    \end{table}
}

In a more recent work (\cite{Collaboration17}), a combination of five independent datasets namely, CMB + DES + BAO + BBN + (SPTpol \cite{Henning17}) along with constraints from Sh0ES \citep{Riess16} and H0liCOW \citep{Bonvin17} was implemented to quote a value of $H_0 = 69.1^{+0.4}_{-0.6} $ \text{km/s Mpc$^{-1} $}, which we are in a very good agreement with. Also, our estimate on $H_0$ which is derived by the inclusion of OHD dataset is very consistent with $H_0 = 71.75 \pm 3.05$ \text{km/s Mpc$^{-1} $} quoted in \cite{Wang17a}, which was obtained using the galaxy-clustering BAO data alone (see also \cite{Cheng15, Chen17, Putten17d, Evslin17}, for constraints on $H_0$ from ``low-redshift'' data). As an attempt to resolve the tension between the P16 and R16 $H_0$ estimates, \cite{DiValentino16} have considered a 12-parameter extended CMB analysis with R16 prior together with an older BAO dataset or with the JLA data \cite{DiValentino16}. This approach clearly resolves the $H_0$ tension and provides values for EOS parameter $w$ consistent with our result. This suggests that for the $w$CDM model the low-redshift analysis performed here is perfectly consistent with the Planck data.

\begin{figure}[ht]
\includegraphics[width=0.45\textwidth]{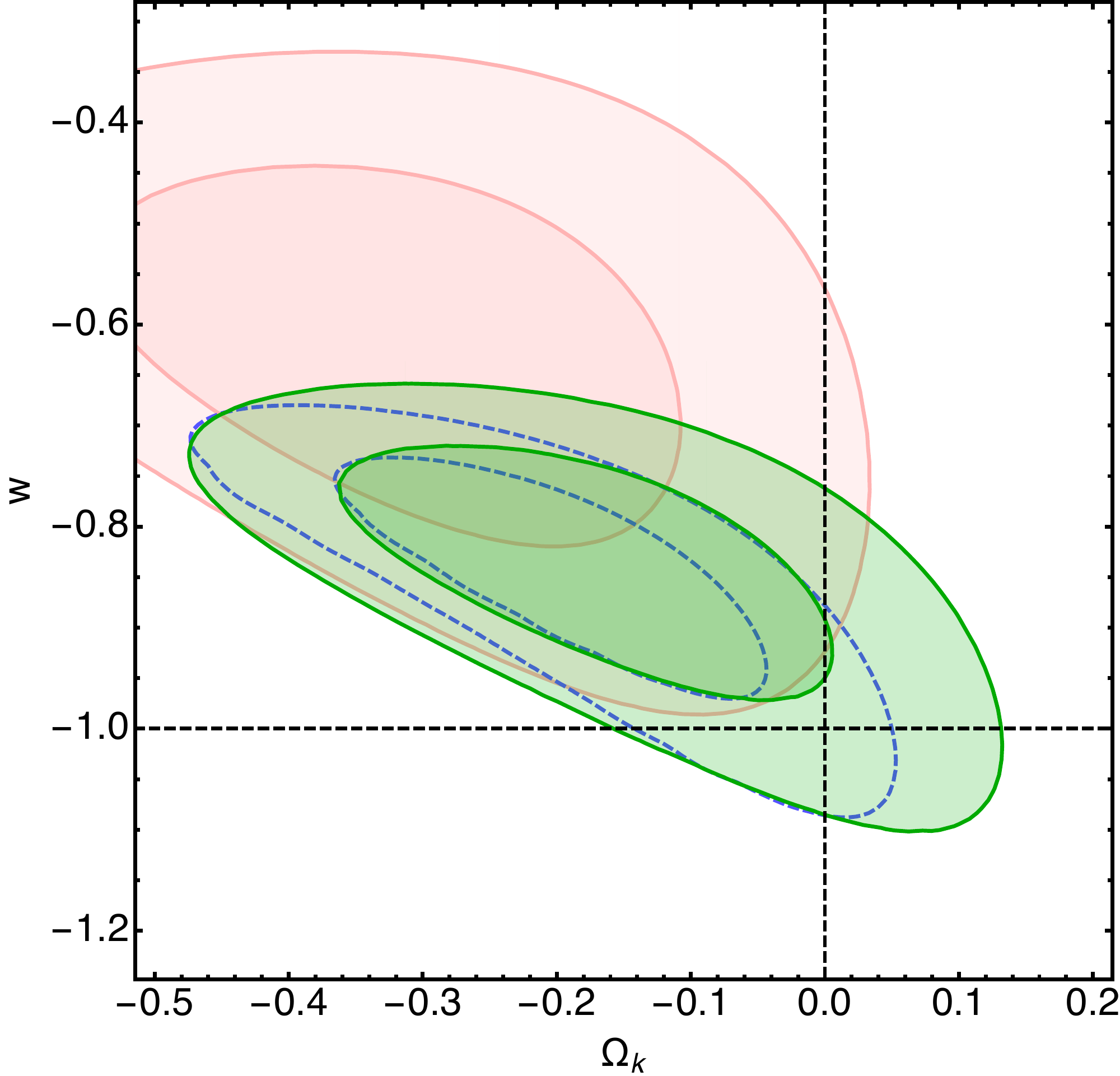}
\hspace{0.2in}
\includegraphics[width=0.45\textwidth]{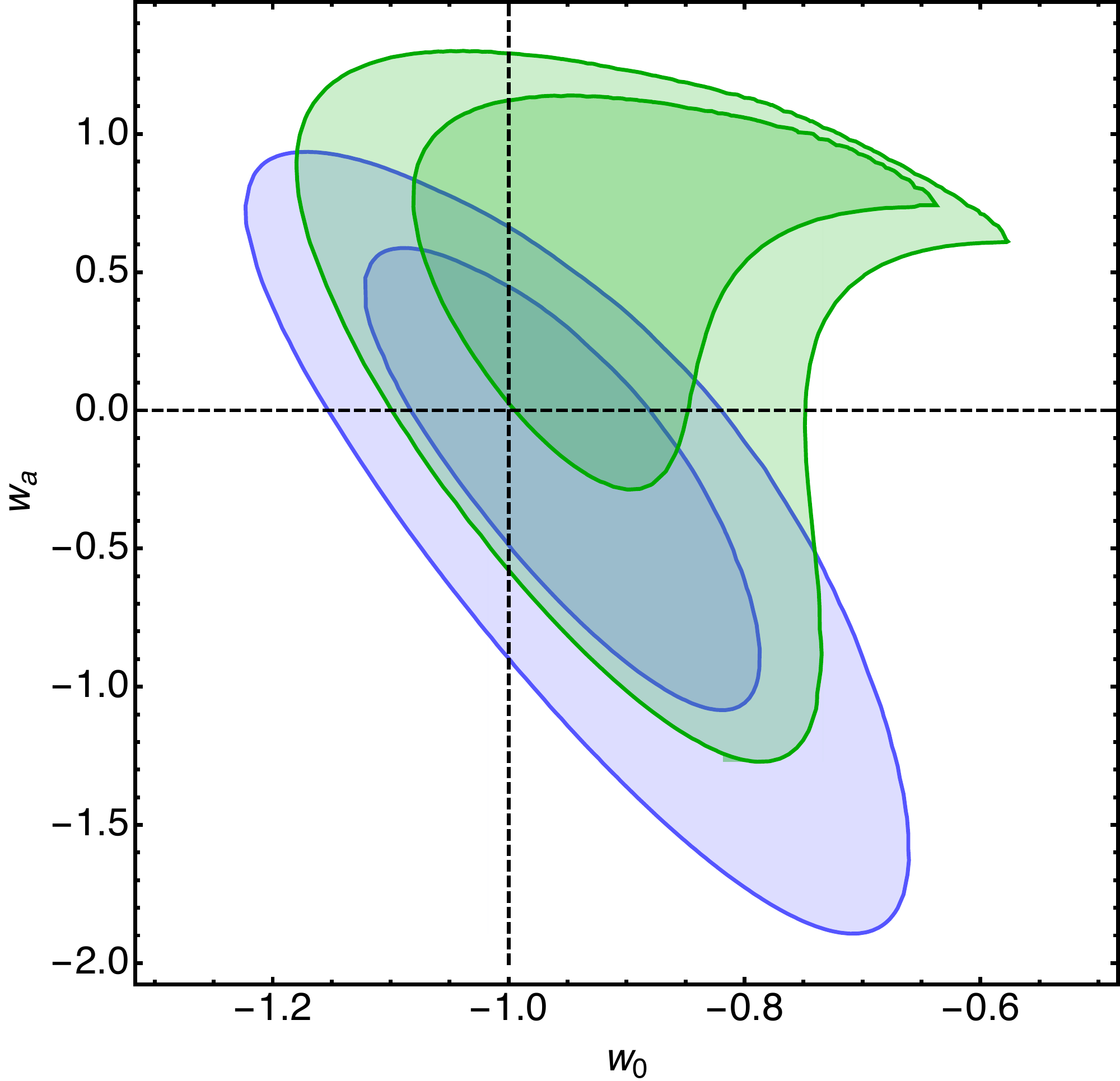}
\caption{\textit{Left panel}: Here we show the 68\% and 95\% confidence level contours obtained for the $kw$CDM model using $D_{M}\&H(z)$+SN+OHD data, with(blue) and without(green) the approximate formula A15. The intersection of the dashed lines ($0, -1$) correspond to the standard model. The red contours correspond to the $kw$CDM fit using the $D_M\&H$ data alone. \textit{Right Panel}: We show the 68\% and 95\% confidence level contours obtained for the $w_0w_a$CDM model using $D_{M}\&H(z)$+SN+OHD data, with(blue) and without(green) the approximate formula A15. The intersection of the dashed lines ($-1, 0$) correspond to the standard model. {Here we show the contours only for the two additional parameters of each model in comparison to standard $\Lambda$CDM model. Please refer to \Cref{tab:joint} for the constraints of all the  parameters.}}
\label{fig:con5}
\end{figure}

We find that an assumption of the approximate formula (A15) in the $w_0w_a$CDM model tends to improve the agreement with the standard model (see right panel of \Cref{fig:con5}), with $\Omega_m = 0.302 \pm 0.016$ and $H_0 = 67.57^{+1.68}_{-1.70}$ \text{km/s Mpc$^{-1} $}. Clearly, there is a degeneracy between $\Omega_m$ and the DE parameters ($w_0, w_a$) that is removed when the A15 formula is assumed. While in the $kw$CDM model the approximate formulae does not show any significant influence on the constraints from joint analysis. However, the $kw$CDM model is in agreement with the standard model at $1.12\sigma$ from our joint analysis, while using $D_M\&H$ data alone shows that the standard model is in agreement only at $2.2\sigma$ (see \Cref{fig:con5}). The individual priors of $ \Omega_k = 0$ in the $w$CDM model and $w=-1$ in $k\Lambda$CDM model tend to converge their model constraints to $\Lambda$CDM. While, leaving both parameters free shows a marginal deviation from $\Lambda$CDM.

Constraints on the $w_{0}w_{a}$CDM model obtained from our joint analysis tend towards the first quadrant of  \Cref{fig:con5} (i.e., $w_0>-1,w_a>0$) and the standard model is consistent at $1 \sigma$. We find that this deviation is slightly larger at $1.5 \sigma$ using $AP$ data alone. Our results are not in immediate agreement with previous results, as the CMB data seems to disfavour the first quadrant \citep{Ade16}. On the same line, the analysis on $w_{0}w_{a}$CDM model presented in \cite{Joudaki17a}, using CMB and weak lensing data also suggests an exclusion of the first quadrant at $\gtrsim 2 \sigma$ confidence. This tension between CMB and BAO was also discussed by \cite{Bernal16} and \cite{DiValentino17}, both using an older compilation of BAO dataset. According to our result, this tension does not seem to change even when the newer BAO dataset is implemented in the analysis.

In \Cref{tab:AIC} we show the values of the information criteria obtained from the joint analysis and $D_M\&H$ data alone for the four models tested against $\Lambda$CDM. $\Delta$AICc strongly disfavour the extended models for $D_M\&H$ data alone, while $\Delta$BIC strongly penalises them in the joint analysis. We find that $w_0w_a$CDM is disfavoured over $\Lambda$CDM with $\Delta$BIC = 10.9. The standard information criteria like BIC tend to heavily penalise models with extra parameters as the number of data points increase. In fact, \cite{Zhao17} predicted an oscillatory nature for dark energy EOS using model-independent technique. They have quoted a $3.5\sigma$ preference for this dynamical dark energy over $\Lambda$CDM using the Kullback-Leibler divergence, which the bayesian evidence is unable to provide. The CPL parametrisation was shown to be a very good approximation for a wide range of scalar field models and modified gravity scenarios in \cite{Linder15}. However, taking into account the discrepant results from the BAO and the CMB data one could infer that the CPL parametrisation is unable to provide conclusive evidence for dynamical nature of dark energy. This evidence should be more generally inferred for a physically motivated $w(z)$ or through a model independent analysis instead of a simple Taylor expansion around $a = 1$, which is imposed over the range of data $0.285 < a < 1$.

{\renewcommand{\arraystretch}{1.45}%
\begin{table}[ht]
\begin{center}
\label{tab:AIC}
\footnotesize
\vspace{0.2in}
\resizebox{.5\textwidth}{!}{
\begin{tabular}{|ccccc|}
    \hline
Model & $\Delta$AICc & $\Delta$BIC & $\Delta${AICc}$_{D_M\&H}$ & $\Delta${BIC}$_{D_M\&H}$  \\
\hline
$w$CDM & 1.09 & 5.71 & 2.15 & 0.16 \\
$k\Lambda$CDM & 1.20 & 6.61 & 3.58 & 1.60 \\
$kw$CDM & 0.97 & 10.20 & 3.05 & -2.63 \\
$w_0w_a$CDM & 1.67 & 10.90 & 7.42 & 1.74 \\
\hline
\end{tabular}
}
\caption{Comparison of the $\Delta$AICc and $\Delta$BIC criteria for the extended models with $\Lambda$CDM as the reference model. The first two columns correspond to the joint analysis and the last two columns are estimated from $D_M\&H$ data alone.}
\end{center}
\end{table}
}

\subsection{Comment on acceleration}
SNIa observations have been the first to provide an evidence for acceleration \citep{Riess98, Perlmutter99}. In a more recent work this evidence has been questioned \cite{Nielsen15} and further discussed \citep{Rubin16, Haridasu17, Tutusaus17, Dam17}. However, the recent BAO data is now capable of constraining the acceleration with much higher significance. In an earlier work, \cite{Ata17} have quoted an evidence of $6.5 \sigma$ using the BAO data alone.
In the $k\Lambda$CDM scenario, using the $D_M\&H$ data, we find, in agreement with \cite{Ata17}, that the acceleration is significant at $5.8 \sigma$. It is important to note that the evidence for acceleration obtained using the BAO data is coming from the $AP$ component. In fact using $AP$ alone, we find that acceleration is preferred at $6.0 \sigma$. Also, from 9z+Ly-$\alpha$ data we have this evidence at $6.6\sigma$.
On the contrary, the $D_V$ component is incapable of constraining the acceleration, while it constraints $\Omega_m$ extremely well (see \Cref{fig:con2}). The evidence for the late-time acceleration increases when a joint analysis is performed: in the $k\Lambda$CDM scenario, we obtain a significance of $8.4 \sigma$. On the other hand, BAO data unlike SNIa is unable to provide a significant evidence for acceleration in the $w$CDM model (see \Cref{fig:con2}).

\begin{figure}[ht]
    \centering
\includegraphics[width=0.45\textwidth]{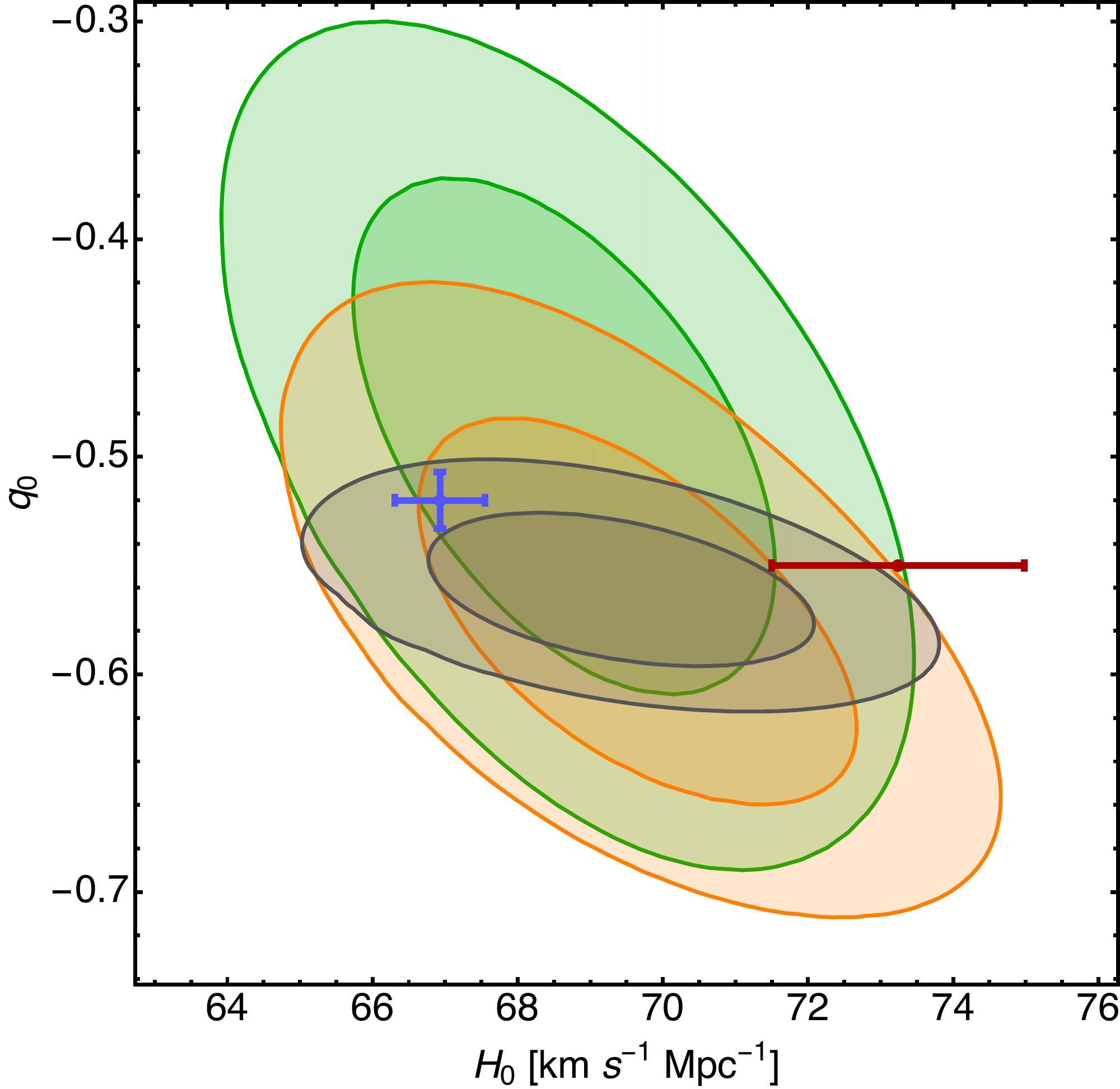}
\caption{Here we show the $q_0$ vs $H_0$ plots for three different models. The green, orange and grey contours are shown for $w$CDM, $k\Lambda$CDM and $\Lambda$CDM, respectively. All contours are made using $D_M\&H$+SN+OHD. In blue we show the Planck constraints \cite{Collaboration16b} and Riess $H_0$ in red.}
\label{fig:con6}
\end{figure}

In \Cref{fig:con6} we show the $H_0-q_0$ plane, where the significance of acceleration and the $H_0$ estimates can be compared simultaneously. We find that all the models predict very consistent $H_0$ and $q_0$ values, while $w$CDM model shows the least evidence for an accelerated scenario.

\subsection{Analysis on mock data}

\begin{figure}[ht]
\includegraphics[width=0.45\textwidth]{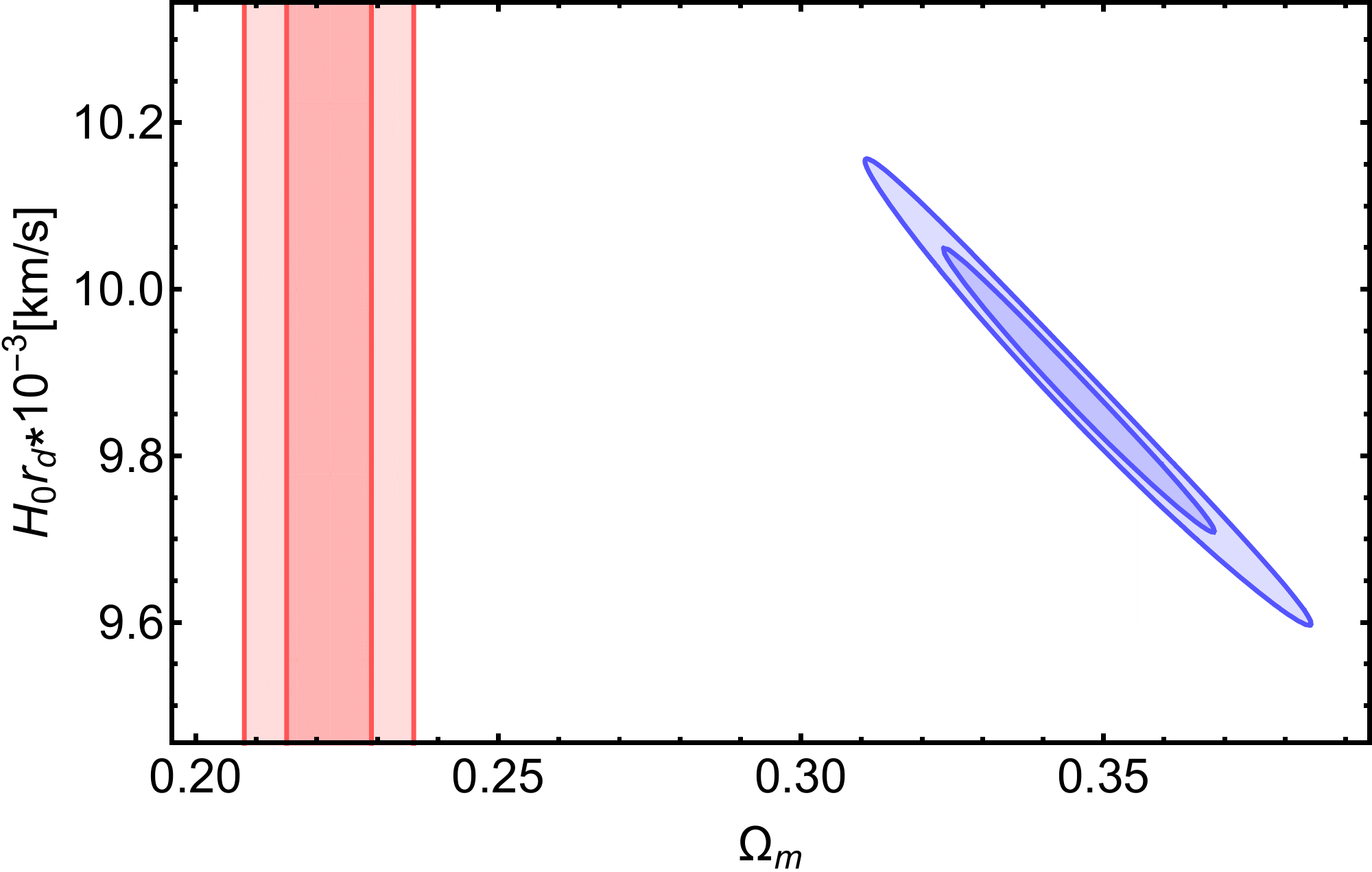}
\hspace{0.1in}
\includegraphics[width=0.45\textwidth]{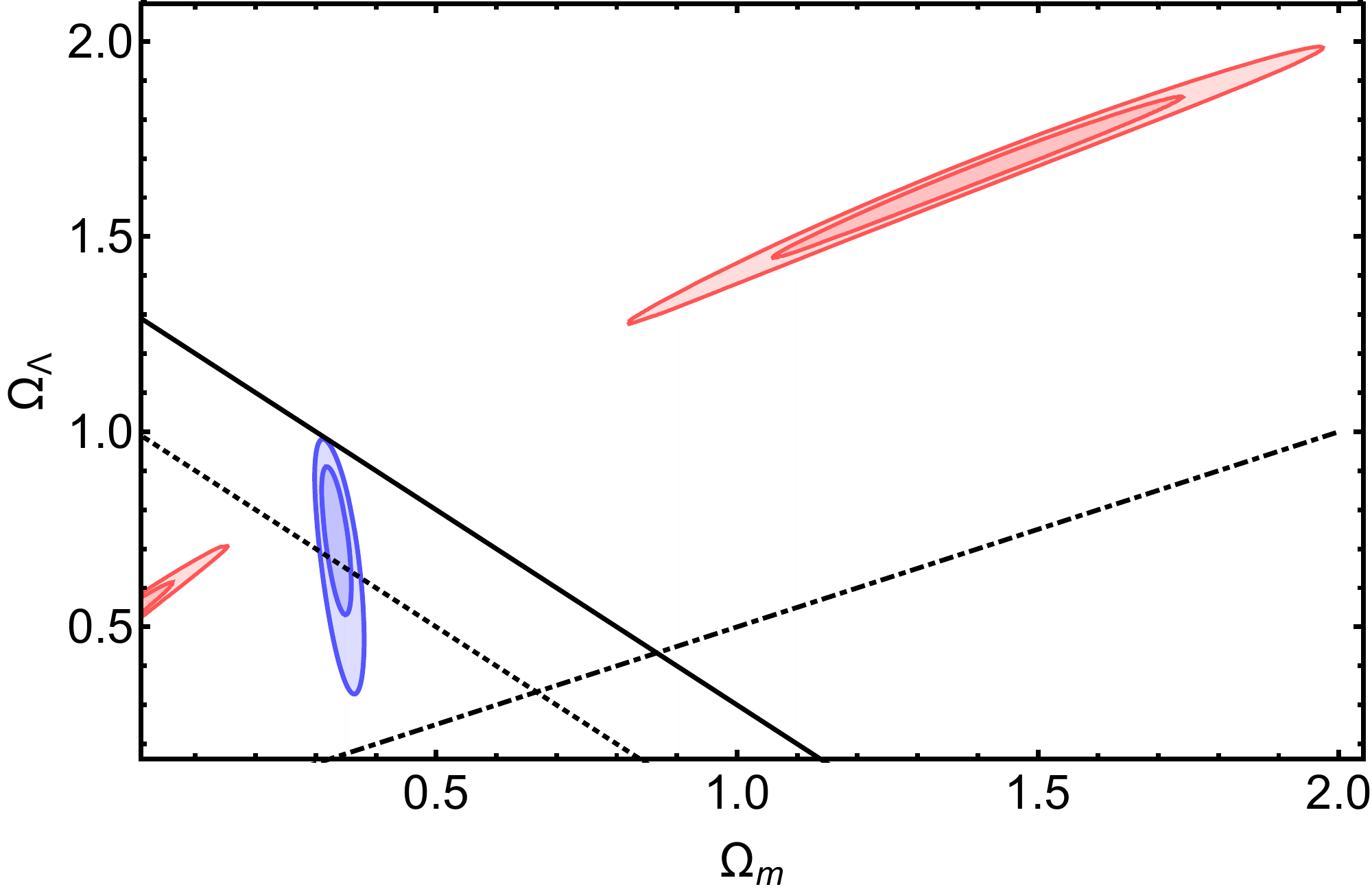}
\vspace{0.1in}
\includegraphics[width=0.45\textwidth]{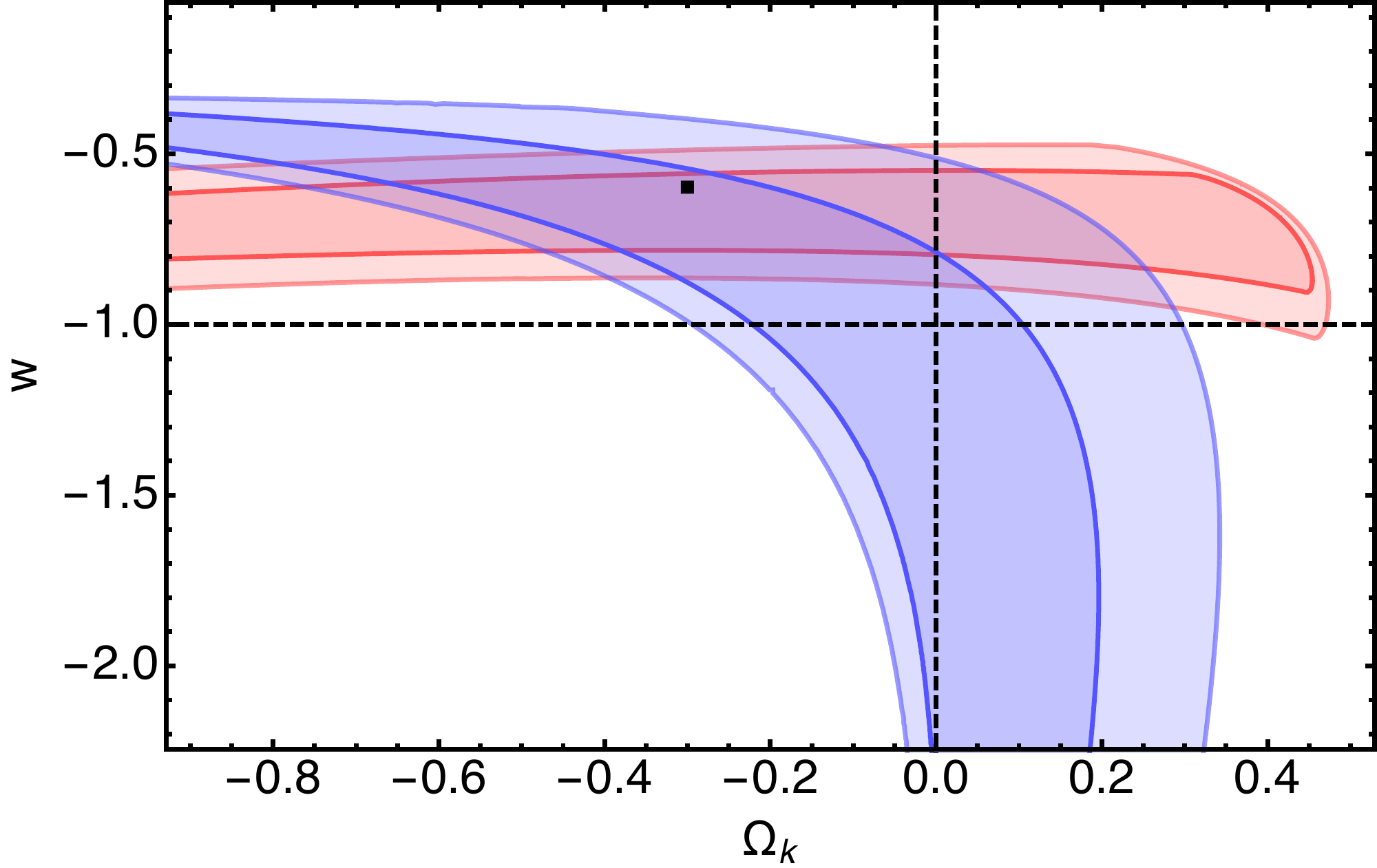}
\caption{In the top-left panel we show the constraints obtained for the $\Lambda$CDM model from our mock dataset using the $D_V$ (blue) and the $AP$ (red) components separately. Similarly, we show the constraints in $k\Lambda$CDM (middle panel) and $kw$CDM (lower panel) scenario. The solid line in the middle panel corresponds to curvature of the model of the mock dataset ($\Omega_k = -0.3$). The dotted line in the middle panel corresponds to flat model, the dot-dashed line marks the transition between the accelerated and non-accelerated regimes. In bottom panel the square marks the ``true'' model around which the mock data has been constructed and the dashed lines show the standard model.}
\label{fig:conmock}
\end{figure}

We implement a mock dataset with Euclid-like (see Table VI. of \cite{Font-Ribera13}) precision built around our best-fit $kw$CDM model  ($\Omega_m$ = 0.3, $\Omega_{\Lambda}$ = 1.0, $w$ = -0.6, $H_0r_d = 9610$ km/s) obtained using $D_M\&H$ data alone. {The mock data for $D_M(z)$ and $H(z)$ observables are obtained from a Gaussian-random distribution around the assumed model with the prescribed relative precision and a constant correlation coefficient of 0.4 at each redshift (see section 1 of \cite{Font-Ribera13}), in the range $0.65 \leq z \geq 2.05$. These $D_M\&H$ measurements were later transformed into the $D_V\&AP$ measurements.} We fit the $AP$ and $D_V$ components of the mock data separately to emphasise the role of their disagreement for model selection. When we fit these two components to $\Lambda$CDM model, the corresponding values of $\Omega_m$ are in a strong tension, at $6.4 \sigma$ (see top-left panel of \Cref{fig:conmock}). Likewise, fitting $k\Lambda$CDM model also shows a strong disagreement as can be seen in the top-right panel of \Cref{fig:conmock}. As expected, once we use the $kw$CDM model, we find a good agreement in the constraints obtained from the $AP$ and $D_V$ components both of which contain the ``true'' model (see bottom panel of \Cref{fig:conmock}). Note, that the $w$ parameter is better constrained by $AP$ rather than by $D_V$.

The ability of the $AP$ and $D_V$ data to provide model selection can also be complemented by the standard information criteria. In fact, the latter indicates a clear preference for the ``true'' model (see  \Cref{tab:AICmock}). However, when the ``true'' model is unknown, even an insignificant evidence for model selection based on standard information criteria can be increased by comparing the $AP$ and $D_V$ constraints. For example, according to standard information criteria, $w$CDM and $w_0w_a$CDM models perform as well as $\Lambda$CDM, even if none of them is the ``true'' model. The $k\Lambda$CDM model is clearly preferred over the previous three models with $\Delta \textrm{BIC} \sim 50$. However, the top-right panel shows that the $AP$ and $D_V$ strongly disagree. Finally, the $kw$CDM is performing better than $k\Lambda$CDM, both using the $\Delta  \textrm{BIC}  \sim 15$, and $AP$ and $D_V$. So, an inspection of the $D_V$ and $AP$ constraints can provide a very useful guideline for model selection even without knowing the true model, as both these components are obtained from same dataset and do not have different systematics.

{\renewcommand{\arraystretch}{1.45}%
\begin{table}[ht]
\begin{center}
\label{tab:AICmock}
\footnotesize
\vspace{0.2in}
\resizebox{.35\textwidth}{!}{
\begin{tabular}{|ccc|}
    \hline
Model & $\Delta${AICc}$_{D_M\&H}$ & $\Delta${BIC}$_{D_M\&H}$  \\
\hline
$w$CDM &  0.68 & 1.60\\
$k\Lambda$CDM & -55.10 & -54.17\\
$kw$CDM & -69.42 & -67.77\\
$w_0w_a$CDM & 2.70 & 4.34\\
\hline
\end{tabular}
}
\caption{Comparison of the $\Delta$AICc and $\Delta$BIC criteria for the extended models with $\Lambda$CDM as the reference model.}
\end{center}
\end{table}

\section{Conclusions}
\label{sec:con}
In this work we have explored how different components of the BAO data constrain cosmological parameters. We find that the isotropic $D_V$ and the anisotropic $AP$ components can provide different constraints when used separately. We find a least possible tension of $1.5 \sigma$ for the $\Omega_m$ values estimated in the $\Lambda$CDM scenario, which increases to $2.3 \sigma$ when the 9z data is used. Although, the current discrepancies are not very significant, similar comparison can be utilised to falsify cosmological models with more precise data to come from future experiments such as EUCLID \citep{Laureijs11} and DESI \citep{Collaboration16, Collaboration16a}. Such a method could be utilised to circumvent the problem of possible systematics in the data. We also present an analysis of a mock scenario with Euclid-like precision \cite{Font-Ribera13} in which we show how these conclusions can be useful to derive significant inferences from forthcoming data.

From our joint analysis, we find $H_0 = 69.41 \pm 1.76$ \text{km/s Mpc$^{-1} $}, which is now consistent with both Planck and R16 at $1.33 \sigma$ and $1.55 \sigma$, respectively. Our findings also very well agree with the most stringent constraint of $69.1^{+0.4}_{-0.6}$  \text{km/s Mpc$^{-1} $} quoted in \cite{Collaboration17}. However, in the $\Lambda$CDM scenario, the three independent $H_0$ estimates from ``low-redshift'' data in this work, ``high-redshift'' P16 and local R16, which are at a total inconsistency of $\textrm{IOI} = 4.19$, increase the difficulty to have a concordance value for $H_0$. Using BAO data in the $D_V$ observable alone can give rise to strict constraints on $\Omega_m$, but not on the cosmic dynamics. This could give the impression that the BAO data is incapable of constraining acceleration. On the contrary, $AP$ provides strong constraints on the acceleration. In fact, the evidence for acceleration is now very strong, at $\sim 5.8\sigma$ from the BAO data alone, and reconfirms the findings from the SNIa data. Using SNIa, OHD and BAO data this evidence is found to be $8.4\sigma$.

In the $kw$CDM scenario, using $D_M\&H$ data alone, we find a mild deviation of $2.2\sigma$ from the standard model. Also in this case, with a $\Delta$BIC = -2.34 the $kw$CDM model is preferred over $\Lambda$CDM. Using the CPL parameterisation we find no evidence for dynamical nature of dark energy from our joint analysis. Although, the BAO data is now providing much better constraints, the dynamical nature of dark energy still eludes in the standard methods for model selection. We also compared the constraints obtained with and without using the A15 approximate formula for the estimation of $r_d$ in 2-parameter extended models. We find a mild difference in the $w_0w_a$CDM model in this comparison.

\section*{Acknowledgements}
B.S.H, V.L and N.V. acknowledge financial support by ASI Grant No. 2016-24-H.0.

    \bibliographystyle{JHEP}
    \bibliography{bao}

\end{document}